# Scale-free collaboration networks: An author name disambiguation perspective

Jinseok Kim

Institute for Research on Innovation & Science
Survey Research Center, Institute for Social Research, University of Michigan
330 Packard Street, Ann Arbor, MI U.S.A. 48104-2910
jinseokk@umich.edu; jinseok.academic@gmail.com

Abstract

Several studies have found that collaboration networks are scale-free, proposing that such networks can be modeled by specific network evolution mechanisms like preferential attachment. This study argues that collaboration networks can look more or less scale-free depending on the methods for resolving author name ambiguity in bibliographic data. Analyzing networks constructed from multiple datasets containing 3.4M ~ 9.6M publication records, this study shows that collaboration networks in which author names are disambiguated by the commonly used heuristic, i.e., forename-initial-based name matching, tend to produce degree distributions better fitted to power-law slopes with the typical scaling parameter ($2 < α < 3$) than networks disambiguated by more accurate algorithm-based methods. Such tendency is observed across collaboration networks generated under various conditions such as cumulative years, 5- & 1-year sliding windows, and random sampling, and through simulation, found to arise due mainly to artefactual entities created by inaccurate disambiguation. This cautionary study calls for special attention from scholars analyzing network data in which entities such as people, organization, and gene can be merged or split by improper disambiguation.

Introduction

A network is called "scale-free" if its node degree distribution follows a power-law pattern of $x^{-\alpha}$, where $x$ is a node degree and $\alpha$ is a scaling parameter (Barabási & Albert, 1999). Scale-free networks have attracted huge scholarly attention due mainly to the implication that complex networks can be modeled by generic principles (Keller, 2005). Until recently, scholars across domains have reported observations of scale-free networks and proposed diverse mechanisms generating such a universal pattern (e.g., Barabási et al., 2002; Pastor-Satorras & Vespignani, 2001).

Among many types of networks, scientific collaboration networks have been confirmed to exhibit scale-free-ness (e.g., Barabási et al., 2002; Milojević, 2010a; Newman, 2001). In a collaboration network, authors are represented by nodes that are connected by edges if two authors appear together in a paper's byline. Conventionally, only the existence of coauthoring relationship between a pair of authors is considered for scale-free network analyses, ignoring collaboration frequency. This means that a node degree in a collaboration network corresponds to the number of distinct coauthors who have ever collaborated with an author represented by the node. In several studies, degree distributions in collaboration networks have been found to follow a power-law: a few authors have large numbers of coauthors while many others have small numbers of coauthors, and this skewness of coauthor distribution fits approximately into a pattern of $x^{-\alpha}$ and, sometimes, across a limited range of $x$ values.

Serving as evidence of scale-free social networks, the aggregated findings of scale-free-ness in collaboration networks have formed an important basis of various efforts to model human interaction patterns besides physical, technical, and biological complex networks (Keller, 2005). Some scholars have, however, reported that degree distributions in collaboration networks do not follow a power-law (Franceschet, 2011; Grossman, 2002; Moody, 2004; Newman, 2004). In addition, several others have noted that scale-free collaboration networks might result from bibliographic data compromised by author name ambiguity (Fegley & Torvik, 2013; Kim & Diesner, 2015). This study takes the latter data-quality approach to understanding scale-free networks.

In bibliographic data, an author is usually represented by an alphabetical string, which can lead to name ambiguity. For example, two distinct authors who have the same names (e.g., two "Charles Brown"s) can be misrepresented as one if we identify authors by their names, which is called "merging of entities." Another ambiguous case would be an author who uses different name variants across papers (e.g., Charles Brown, Charles C. Brown, and Charlie Brown), causing the work of the author to be attributed to multiple other authors, called "splitting of entities."

To address this ambiguity problem, many scale-free collaboration networks have been constructed under the assumption that two names that match on forename initials and surname refer to the same author. This initial-based author matching can produce disambiguation errors by mismatching two distinct authors who share the name initials (e.g., Charles Brown and Clarke Brown) or mistakenly regarding two names (e.g., Charles Brown and Charles C. Brown) of an author as belonging to different authors. Scale-free collaboration network studies using this initial-based heuristic have well acknowledged the misidentification problem but argued that the initial-matching-induced errors would not change "much" knowledge discovered from ambiguous bibliographic data (Barabási et al., 2002; Newman, 2001).

To counter-argue the negligible impact of author name ambiguity on collaboration networks, this study shows that scale-free-ness of collaboration networks can be affected by artefactual nodal entities created by ambiguous author names. In doing so, this study uses three large-scale bibliographic datasets to construct collaboration networks in which author names are disambiguated by three different methods – all forename initials plus surname, a first forename initial plus surname, and algorithmic disambiguation. Then, a power-law fitting test is conducted for degree distributions of collaboration networks generated under various conditions such as 5- & 1-year sliding windows, cumulative years, and random selection of paper records. In addition, how merged or split author entities are related to the rise of scale-free networks is simulated with incremental changes in disambiguation errors.

## Methodology

### Datasets

This paper analyzes collaboration networks constructed from three large-scale scholarly datasets covering biomedicine, physics, and computer science. This selection represents academic fields that have been frequently studied by researchers for scale-free networks as well as bibliometrics in general.

MEDLINE: Maintained by the U.S. National Library of Medicine, this dataset contains almost 24M publication records published in biomedicine-related journals worldwide. The 2016 baseline data were downloaded in XML format[1]. As MEDLINE author names are not disambiguated, the Author-ity data containing disambiguated MEDLINE author names for the 1991 ~ 2009 period (Torvik & Smalheiser, 2009; Torvik, Weeber, Swanson, & Smalheiser, 2005)[2] were obtained. Author names in Author-ity are disambiguated through machine learning consisting of two steps: (1) pairwise similarity comparison on name strings and metadata information such as journal name, title, affiliation, and MeSH term and (2) a maximum likelihood based, agglomerative clustering algorithm (Torvik & Smalheiser, 2009). Author names in MEDLINE were assigned Author-ity IDs through matching record instances in MEDLINE and Author-ity using PMIDs (paper identifiers in MEDLINE) and an author name's position in a paper's byline. This matching resulted in a total of 9.6M paper records containing 39.3M author name instances of 6.3M distinct authors[3].

MAG: Microsoft Academic Graph (MAG) is a bibliographic dataset of publication records crawled by Microsoft's search engine[4]. The downloaded 2016 version contains author identifiers assigned by a crude level disambiguation algorithm conducted for a release purpose (Sinha et al., 2015). The details of "various best-effort algorithms" used for disambiguating names in MAG are not disclosed possibly because they are proprietary. From the bulk data, a subset of journal papers that are published in the broad area of physics during the 1991~2015 period was selected[5]. This selection produced a total of approximately 4.3M paper records that contain 4.9M unique author identifiers associated with 19.2M author name instances

DBLP: The Digital Bibliography & Library Project (DBLP) is a digital library curating paper records published in computer science (Ley, 2002). The whole DBLP data are released monthly[6]. The 2017

---

[1] https://www.nlm.nih.gov/databases/download/pubmed_medline.html
[2] https://databank.illinois.edu/datasets/IDB-4222651
[3] To reduce the distortion of degree distribution by hyper-authorship, paper records with many authors were excluded from each dataset. Specifically, when decreasingly ordered by the number of authors per paper, top 1% of all papers in each dataset were omitted before analysis. The threshold of exclusion was $9 \leq$ in DBLP, $16 \leq$ in MAG, and $13 \leq$ in MEDLINE. For comparison, power-law fitting procedure was conducted on datasets with papers authored by more than 100 authors excluded, which are reported in Appendix A.
[4] https://www.microsoft.com/en-us/research/project/microsoft-academic-graph/
[5] MAG sub-categories of physics include: acoustics, astronomy, astrophysics, atomic physics, classical mechanics, condensed matter physics, mechanics, nuclear physics, optics, optoelectronics, particle physics, quantum electrodynamics, quantum mechanics, statistical physics, theoretical physics, and thermodynamics.
[6] http://dblp.org/xml/release/

September version was used in this study. DBLP author names are disambiguated by the combination of algorithms and human curation: (1) author names are grouped first by name string matching and coauthor similarity, (2) merged or split author names from (1) are corrected by community detection algorithm and (3) suspicious cases from (2) are manually checked by the DBLP team and users (Müller, Reitz, & Roy, 2017). Excluding records of books and dissertations, almost 3.4M paper records published during the 1991~ 2016 period were used to construct collaboration networks of 1.8M unique authors associated with approximately 10M name instances.

Pre-processing: Following the dominant practice of scale-free collaboration network studies, author names in each dataset were changed into the format of all forename initials and a full surname(s) (e.g., Charles C. Brown → C. C. Brown). This method is mentioned as AINI hereafter. In addition, each name was converted into the simplest format, i.e., the first forename initial followed by a full surname(s) (e.g., Charles C. Brown → C. Brown). This first-initial method (FINI hereafter) has been used as a standard author reference format in academia for a long time (Garfield, 1969) as well as a method for disambiguating author names in collaboration network research (e.g., Liben-Nowell & Kleinberg, 2007; Newman, 2001).

Disambiguation Accuracy

*Merging versus Splitting*: Author name ambiguity can affect a degree distribution by merging and splitting entities in collaboration networks (Kim, 2017). In Figure 1, for example, two author nodes (Charles C. Brown and C. C. Brown) in Network A are connected to two alters (coauthors): i.e., each author has a degree of 2. According to AINI, these two nodes are consolidated into one (C. C. Brown in Network B), decreasing the number of nodes to one. As a result of this merging, two sets of coauthors ([C1, C2] and [C3, C4]) in Network A are attached to the merged author entity (C. C. Brown) in Network B, who now has a degree of 4. If many authors share the same initialized forename and full surname, the merging can produce an artificial entity with a large number of coauthors (i.e., a large degree). The reverse of this merging shows the impact of entity splitting. The node of a single author (C. C. Brown) in Network B may be divided into two nodes by an algorithmic decision that Charles C. Brown and C. C. Brown refer to different persons, resulting in fragmented node degrees.

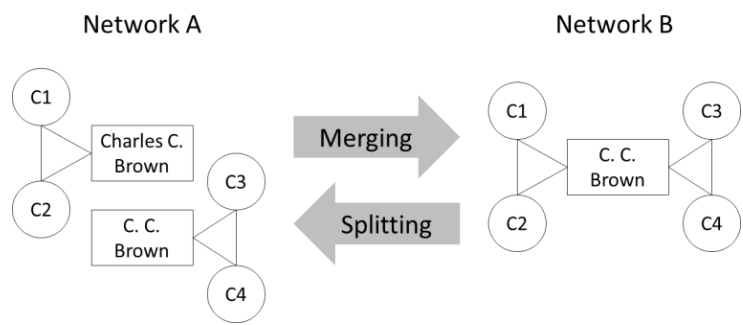

*Figure 1: An Illustration of Merging and Splitting of Author Entities by Ambiguous Names*

*Labeled Truth Data*: To find out how much each dataset is susceptible to merging and splitting errors by different disambiguation methods, this study uses ORCID author profiles to construct ground truth for evaluating disambiguation accuracy. The ORCID is an information system of scholarly profiles managed by authors who register their publication records and auxiliary information such as affiliation and emails (Haak, Fenner, Paglione, Pentz, & Ratner, 2012). The whole ORCID dataset containing nearly 3.5M author profiles as of 2017 October was obtained[7]. Then, each paper's title in this study's data was compared to publication records of ORCID-registered authors. Specifically, this study pre-processed titles both in ORCID and bibliographic data by (1) converting special characters into ASCII, (2) changing alphabet characters in lower-case, (3) removing mechanics (e.g., period), (4) excluding stop-words (e.g., "the"), and deleting spaces. Then, titles that appear twice or more in bibliographic data were excluded from matching to avoid duplicate matches. In ORCID, the same title can appear multiple times because authors of a paper claim their authorship individually in their own ORCID profiles. So, an author name in a paper matched to one of publication records under an ORCID profile was assigned the ORCID ID of the profile owner if the author name matches with the owner's name in a full name format. This matching process was conducted for paper records in MEDLINE and MAG. In DBLP, an author name is already associated with an ORCID ID, if available. So, this study used the list of author name-ORCID ID pairs as recorded in DBLP, following Kim (2018). Table 1 summarizes the numbers of author name instances in three datasets matched to ORCID IDs and the numbers of unique authors identified by ORCID IDs in comparison with those by algorithmic, AINI, and FINI methods.

*Table 1: Summary of Record Matching Results between Three Datasets and ORCID Author Profiles*

| Data | Number of Instances | Number of Distinct Authors | | | |
|---|---|---|---|---|---|
| | | ORCID | Algorithmic | AINI | FINI |
| MEDLINE-ORCID | 940,410 | 130,712 | 137,262 | 115,122 | 109,410 |
| MAG-ORCID | 770,534 | 136,866 | 178,559 | 114,443 | 108,728 |
| DBLP-ORCID | 664,472 | 103,335 | 105,217 | 93,569 | 82,150 |

*Accuracy Measurement*: According to the table, the numbers of authors detected by initial-based disambiguation (AINI and FINI) were smaller than those by algorithmic methods, possibly due to the merging effect as described in Figure 1. However, how many authors are merged or split is unknown in the table. To measure how often merging and splitting happens, the ratios of authors who are merged, split, or both merged and split in each dataset were calculated (Kim & Diesner, 2016). For this purpose, specifically, all ORCID-linked name instances in each dataset were assigned unique instance IDs. Next, instance IDs belonging to the same ORCID IDs were grouped to form truth clusters (each cluster represent a single author). In the same way, instance IDs belonging to the same authors identified by, e.g., AINI were collected to form test clusters. Then, each truth cluster was checked against the list of test clusters to see whether (1) "all and only" the instance IDs in the target cluster appear in the same test cluster (i.e., correctly disambiguated), (2) any instance ID that does not belong to the target cluster appears together in a test cluster with any of the target cluster's instance IDs (i.e., merged), (3) any instance ID of the target cluster appears in other test clusters (i.e., split), or (4) both (3) and (4) happens (i.e., merged & split). As such, this cluster-based error checking enables us to decide each truth cluster (= a true author) to be error-free, merged, split, or merged and split by a disambiguation method[8].

*Accuracy Test Results*: Table 2 reports the ratios of misidentified authors by three disambiguation methods – algorithmic, AINI, and FINI – tested on the truth datasets (from Table 1) of author name instances linked to ORCID IDs. Ratios of merged authors by algorithms were less than 2% across three ORICD-linked datasets, indicating that algorithm-based disambiguation could distinguish very well name instances belonging to distinct authors. Meanwhile, ratios of split authors are a little higher than merging ratios but still low (4.54% for MEDLINE-ORCID and 2.63% for DBLP-ORCID). For MAG-ORCID, however, splitting errors (22.42%) were substantial, meaning that the disambiguation algorithm used for MAG failed to find name instances that should belong to distinct authors in many cases[9]. The high level of splitting in MAG-ORCID produced fragmented author entities (as shown in Figure 1), resulting in the larger number of distinct authors (178,559 in Table 1) than the true number of authors (136,866 in Table 1) by ORCID IDs.

---

[7] https://figshare.com/articles/ORCID_Public_Data_File_2017/5479792/1
[8] This is derived from the standard Cluster Recall (i.e., the ratio of truth clusters with no disambiguation error over all truth clusters) combined with the Lumping & Splitting measure in Torvik and Smalheiser (2009).

*Table 2: Disambiguation Errors in Three Datasets Per Disambiguation Method: Values in percentage denote ratios of ORCID authors that are correctly disambiguated (NoError), merged, split, or merged and split by Algorithmic, AINI (all-initials-based), and FINI (first-initial-based) disambiguation methods*

| Error Type | MEDLINE-ORCID | | | MAG-ORCID | | | DBLP-ORCID | | |
|---|---|---|---|---|---|---|---|---|---|
| | Algorithmic | AINI | FINI | Algorithmic | AINI | FINI | Algorithmic | AINI | FINI |
| No Error | 93.71% | 74.64% | 71.69% | 75.70% | 72.88% | 68.65% | 95.59% | 77.91% | 69.04% |
| Merged | 1.62% | 21.79% | 26.10% | 1.29% | 24.47% | 29.30% | 1.74% | 17.98% | 29.20% |
| Split | 4.54% | 2.74% | 1.69% | 22.42% | 1.55% | 1.11% | 2.63% | 3.52% | 1.17% |
| Merged & Split | 0.13% | 0.84% | 0.52% | 0.59% | 1.55% | 0.93% | 0.04% | 0.59% | 0.59% |

Initial-based heuristics for name disambiguation performed very well in reducing splitting errors (1.17% ~ 3.52%). Especially, FINI is shown to be better at collating name instances that should belong to a distinct author than AINI. But this low-level splitting by AINI and FINI was achieved with high merging error ratios ranging from 17.98% to 29.30%. FINI shows higher ratios of merging than AINI because AINI splits authors having different middle forename initials while FINI merges them. For example, C. C. Brown and C. W. Brown are split by AINI but merged by AINI. The high merging ratios by both AINI and FINI means that the initial-based disambiguation combined many distinct authors into artefactual entities, substantially reducing the numbers of distinct authors. In Table 1, for example, the number of distinct authors in MEDLINE-ORCID decreased from 130,712 by ORCID IDs to 115,122 (−11.93%) by AINI and 109,410 (−16.30%) by FINI.

Power-law Fitting Method

A degree distribution of a scale-free network is assumed to follow a power-law pattern defined as $p(x) \approx x^{-\alpha}$, where $x$ is a degree value, $p(x)$ its probability, and $\alpha$ a scaling parameter (also called an exponent) that is constant across $x$ values. Some scholars have tested power-law fitting on all $x$ values in a degree distribution while others on a range of $x$ values that are optimally selected to fit best a power-law slope. A problem is that depending on the choices of the range of target $x$ values and fitting methods, the same degree distribution can be decided to obey a power-law or not (Clauset, Shalizi, & Newman, 2009; Stumpf & Porter, 2012). As there is no consensus on "acceptable" or "legitimate" ranges of $x$ values and this study aims to show how name ambiguity affects the degree distribution shape in collaboration networks, power-law slopes are tested on all *x* values. For comparison, however, the results of power-law fitting on *x* ≥ minimum are presented in Appendix B.

---

[9] This high splitting may be a result of an algorithmic decision by the MAG data team who clarified that, for an academic release purpose, a basic level of disambiguation was conducted for MAG. In addition, it seems that disambiguation design for DBLP and disambiguated MELINE (i.e., Author-ity) aimed at less merging (≈ high precision) than less splitting (≈ high recall) because merging is more detrimental to bibliometrics and network analysis than splitting (Fegley & Torvik, 2013; Kim, 2018; Müller et al., 2017).

Following the common practice of many scale-free network studies, this paper fits a degree distribution to a power-law by projecting it on log-log-scaled axes, estimating its scaling parameter ($\alpha$) by conducting a least-squares linear regression on the distribution plot with an R-squared goodness-of-fit ($R^2$) calculated. Specifically, a node degree distribution of a given network is converted into the complementary cumulative density function (CDF), where the ratio of the number of nodes with *x* degree or more over the total number of nodes is calculated for each *x* value. Then, data points of the function are depicted on doubly logarithmic axes, where *x*-axis denotes degrees (*x*) and *y*-axis denotes ratios of nodes with *x* or more. Figure 2, for example, shows the CDF plot of degree distribution from 9.6M MEDLINE records where author names are disambiguated by AINI. According to the figure, authors who have 10 or more coauthors constitute 44.45% (= 0.4445 on *y*-axis) of all authors. The estimated *α* is 2.6834 by a least-squares (LS) regression with $R^2$ = 0.9802. The fitted power-law slope is represented by a solid line[10].

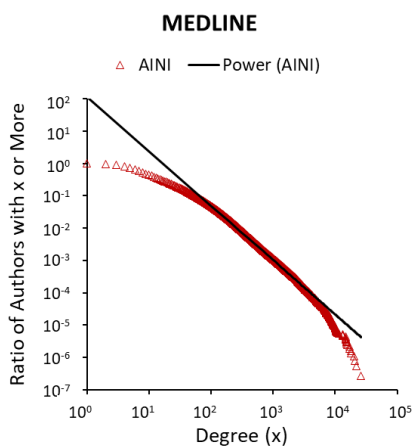

*Figure 2: An Illustration of Power-Law Fitting Test on MEDLINE Disambiguated by All-Initials-Based Disambiguation (AINI)*

Results

Figure 3 shows the degree distributions of three datasets plotted on doubly logarithmic axes. In the figure, degree distributions from algorithmically disambiguated datasets are represented by blue circles, while those from the all-initial-based method (AINI) and the first-initial-based method (FINI) are depicted by red triangles and green crosses, respectively. A common observation across subfigures is that blue circles appear below red triangles and green crosses.

---

[10] The CDF-based estimation of scaling parameters is preferred over the use of probability density function (PDF) because CDF can provide more robust estimation than PDF when distribution tails have fluctuations (Newman, 2005). Tested on power-law obeying synthetic data, however, the CDF-LS method for estimating scaling parameters underperforms than the maximum likelihood estimator (MLE) combined with the Kolmogorov-Smirnov (KS) distance measure (Clauset et al., 2009). Furthermore, some scholars have recommended the use of PDF-based logarithmic binning because cumulative distribution can misrepresent the characteristic of a discrete distribution tail (Milojević, 2010b). As this study aims to show how degree distributions can be affected by author name disambiguation, the commonly used CDF-LS method is believed to suffice to serve the purpose.

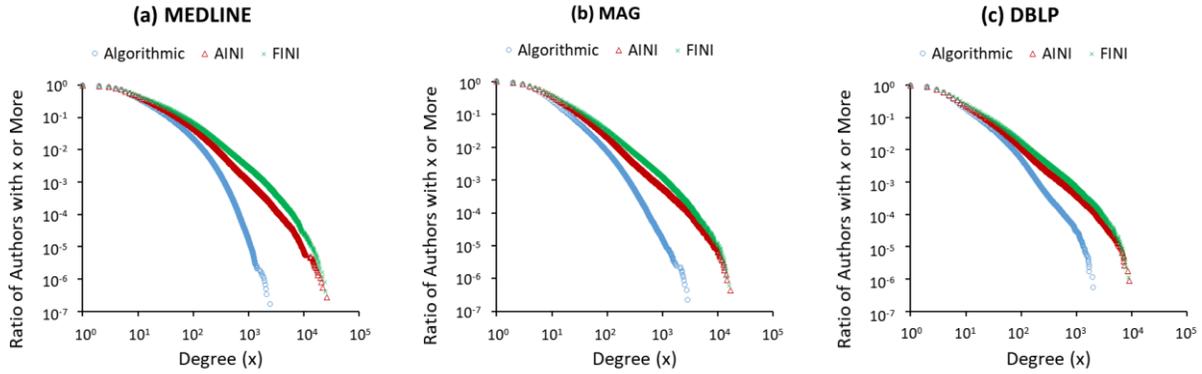

Figure 3: Degree Distribution Plots on Double Logarithmic Scales for Three Datasets

This placement pattern can be explained mainly by merging. First, as shown in Table 2, initial-based disambiguation tends to merge author entities into artefactual ones, attaching the coauthors of merged authors to the alloyed entities. For example, the mean degree of authors in algorithmically disambiguated MEDLINE was 16.01 ($SD$ = 31.95), which increased to 26.23 ($SD$ = 105.76) by AINI and 36.04 ($SD$ = 163.98) by FINI. As artefactual entities amalgamated by multiple authors come to have inflated numbers of coauthors (numerator↑) while reducing the numbers of distinct authors (denominator↓), the ratios of authors who have a specific number ($x$ degree) or more of coauthors also increase. This explains why red and green data points were positioned vertically higher than blue ones. For example, in MEDLINE, the ratio of authors with 10 or more coauthors increased from 37.91% by algorithmic disambiguation to 44.45% by AINI to 48.63% by FINI. The combination of these two merging-induced effects – increasing coauthor sizes and decreasing numbers of author entities – by initial-based disambiguation pushed the degree distribution plots from algorithmically disambiguated data (that are less susceptible to merging) toward upper-right corners in the figure. As FINI tends to merge more authors than AINI, data points for FINI appeared higher than those for AINI in each subfigure.

Same data disambiguated by different methods gave rise to degree distributions with different $\alpha$ and $R^2$. Table 3 summarizes the power-law fit test results. The sizes of $\alpha$ by the initial-based disambiguation were smaller (i.e., slopes became less steep) than those by algorithmic methods. This decrease of $\alpha$ is in line with the tendency of upper-right moving distribution plots by AIN and FIN in Figure 3. Especially, the scaling parameters by AINI and FINI were around $\alpha$ = 2.5, falling within the typical 2 < $\alpha$ < 3 range of scale-free networks (Börner, Maru, & Goldstone, 2004; Dorogovtsev & Mendes, 2002). In contrast, the scaling parameters from algorithmically disambiguated networks were outside $\alpha$ = 3 in MEDINE and MAG but very close to $\alpha$ = 3 in DBLP. Another notable observation is that the $R^2$ values by AINI and FINI were higher than those by algorithmic disambiguation. This implies that initial-based disambiguation produced degree distributions better fitted to power-law slopes than algorithm-based disambiguation. AINI tends to produce slightly higher $R^2$ and larger $\alpha$ than FINI.

Table 3: Summary of Power-Law Fit Test for Three Datasets per Disambiguation Method: $\alpha$ = scaling parameter and $R^2$ = R-squared goodness-of-fit

| Data | Algorithmic | | AINI | | FINI | |
|---|---|---|---|---|---|---|
| | $\alpha$ | $R^2$ | $\alpha$ | $R^2$ | $\alpha$ | $R^2$ |
| MEDLINE | 3.5677 | 0.9079 | 2.6834 | 0.9802 | 2.5718 | 0.9659 |
| MAG | 3.3176 | 0.9623 | 2.5537 | 0.9877 | 2.5409 | 0.9753 |
| DBLP | 3.0743 | 0.9750 | 2.5163 | 0.9862 | 2.4910 | 0.9784 |

Cumulative Years

Conducting over-time analyses of collaboration networks, several scholars have showed that power-law degree distribution can emerge from evolving networks and proposed mathematical models such as preferential attachment to explain underlying mechanisms that give rise to such a law-governed distribution pattern (e.g., Barabási et al., 2002; Milojević, 2010a). Following this practice, collaboration networks were constructed from each dataset starting from 1991 and cumulating up to a target year with yearly increments (e.g., networks of 1991 → 1991~1992 → 1991~1993 → 1991~1994, etc.). Then, degree distributions of each cumulative-year network were fitted to power-law obeying slopes. Figure 4 visualizes the results on two dimensional panes where $α$ is denoted on x-axes and $R^2$ on y-axes. The arrow-headed lines along data points show the recency of target years: the tails represent old years and the arrow-heads recent ones.

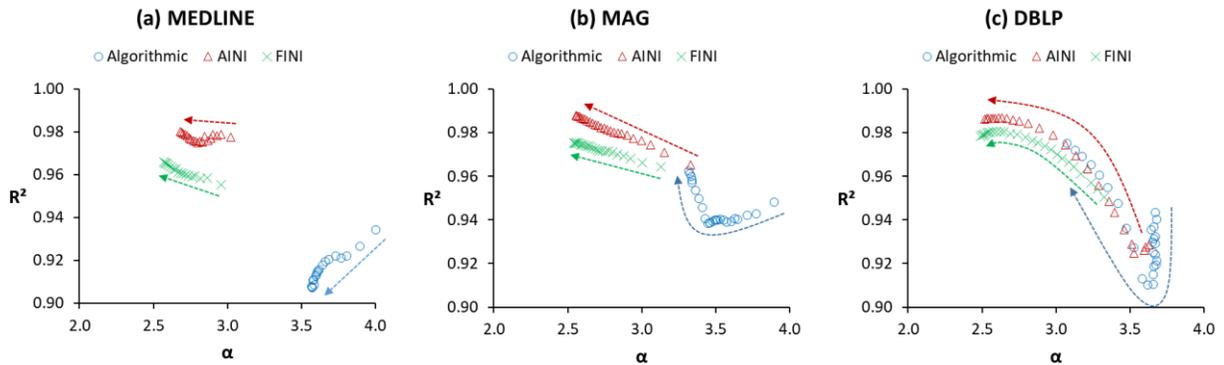

*Figure 4: Trends of Scaling Parameter (α) and R-squared Fit ($R^2$) per Disambiguation Method over Cumulative years: arrow-headed lines represent the recency of target years from old to recent ones*

The figure shows that across three datasets, collaboration networks disambiguated by initial-based disambiguation tended to move toward upper-left corners with decreased $α$ and increased $R^2$. The all-initials-based disambiguation (AINI), commonly used in scale-free collaboration network studies, produced slightly larger $α$ and higher $R^2$ than FINI. This trend is visualized by red triangles positioned above (y-axes) and right side (x-axes) of green crosses. Especially, many data points by initial-based disambiguation moved from right to left on x-axes within the range of 2 < α < 3, densely clustered toward α = 2.5. This implies that, when disambiguated by initial-based disambiguation, collaboration networks in three datasets can be seen to evolve toward scale-free ones with power-law-like distributions, with $R^2$ as close to 0.99 and α that is typical for scale-free networks.

In contrast, the over-time trend of $R^2$ by algorithmic disambiguation is not consistent across datasets, while α kept decreasing. In MEDLINE, blue circles move left-downward: both α and $R^2$ decreased, implying that degree distributions became less and less fitted into power-law slopes. In MAG and DBLP, blue-circled data points formed V-shaped patterns of change over years. This means that degree distributions moved away from power-law slopes in terms of $R^2$ but at some points, began to get closer to them with scaling parameters approaching α = 3.3 (MAG) and 3.0 (DBLP). These observations imply that depending on the choice of cumulative years and data (e.g., the 1991-2016 period for DBLP), likely power-law degree distributions may be observed in algorithmically disambiguated bibliographic data.

5-Year & 1-Year Sliding Windows

Instead of investigating over-time changes of collaboration networks, some studies have analyzed degree distributions of snapshot collaboration networks for a specific period of years ranging from 1 to 20 years (e.g., Börner et al., 2004; Newman, 2001; Wagner & Leydesdorff, 2005). Following this practice, each dataset was divided into subsets of publication records filtered by (1) a sliding 5-year window up to a target year with yearly resolution (e.g., 1987~1991 for the target of 1991, 1988~1992 for the target of 1992, 1989~1993 for the target of1993, etc.)[11] and (2) per year (i.e., a single year window). Then, degree distributions of each network were tested for power-law fit. Figure 5 and Figure 6 shows the results. The arrow-headed lines in both figures show the recency of target years.

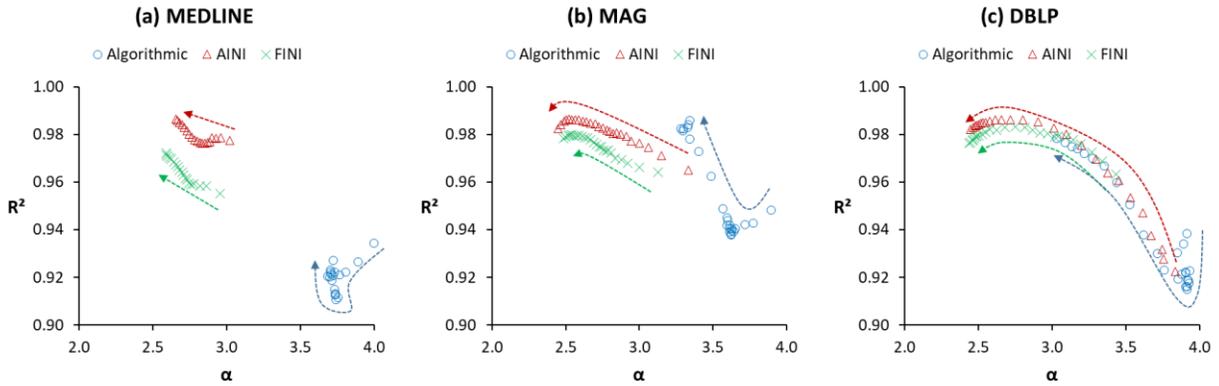

Figure 5: Trends of Scaling Parameter (α) and R-squared Fit ($R^2$) Per Disambiguation Method over 5-year Sliding Window: arrow-headed lines represent the recency of target years from old to recent ones

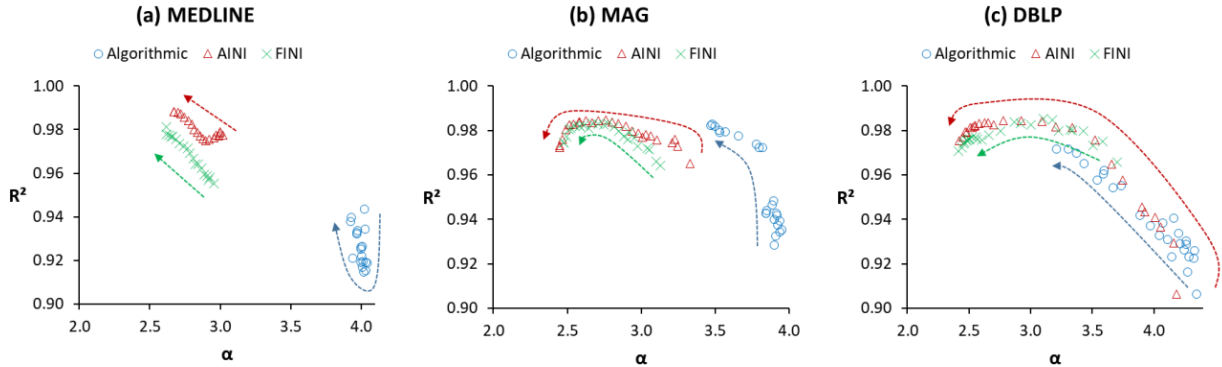

Figure 6: Trends of Scaling Parameter (α) and R-squared Fit ($R^2$) Per Disambiguation Method over 1-year Sliding Window: arrow-headed lines represent the recency of target years from old to recent ones

In both Figure 5 and 6, data points from initial-based disambiguation moved toward upper-left corners: i.e., higher $R^2$ and lower α. Also, data points by the commonly used AINI moved slightly behind FINI ones on *x*-axes (≈ larger α) with higher $R^2$ on *y*-axes. Furthermore, α also fell within the aforesaid typical range (2 < α < 3). In this way, the visualized patterns of moving data points in Figure 5 and 6 are very similar to those in Figure 4 for cumulative years. Degree distributions from algorithmically disambiguated MAG and DBLP in Figure 5 and 6 also showed similar patterns to those for cumulative years in Figure 4. A difference is that although stuck in the lower-right corner, blue data points of MEDLINE in Figure 5 and 6 also formed V-shaped patterns like MAG and DBLP.

---

[11] For MAG and DBLP, 5-year window sliding was performed dating back earlier than 1991. For MEDLINE, however, due to the lack of disambiguated records for before-1991, the sliding window for 1991~1994 was done for part of years: 1991, 1991~1992, 1991~1993, 1991~1994 and then 1991~ 1995, 1992~1996, etc.

Such similarities can be confirmed in Table 4 in which the mean α and mean $R^2$ per disambiguation method are compared across the results based on cumulative years, 5-year sliding window, and one-year window. Across datasets, both mean α and mean $R^2$ are within a few percent of differences between three different time-slicing methods. The 5- and 1-year sliding window tests also show that depending on the choice of 5-year periods and data, algorithmically disambiguated bibliographic data may produce likely power-law degree distributions, with higher α and lower (sometimes, similar to) $R^2$ than those by initial-based disambiguation.

Table 4: Summary of Mean Scaling Parameter (α) and R-squared Fit ($R^2$) per Disambiguation Method for Different Yearly Coverages: standard deviations reported in parentheses

| Data | Year Coverage | Algorithmic | | AINI | | FINI | |
|---|---|---|---|---|---|---|---|
| | | Mean α (SD) | Mean $R^2$ (SD) | Mean α (SD) | Mean $R^2$ (SD) | Mean α (SD) | Mean $R^2$ (SD) |
| MEDLINE | Cumulative Years | 3.6668 (0.1184) | 0.9161 (0.0070) | 2.8112 (0.0942) | 0.9774 (0.0016) | 2.6923 (0.1026) | 0.9616 (0.0029) |
| | 5-Year Window | 3.7533 (0.0733) | 0.9198 (0.0059) | 2.8059 (0.1021) | 0.9798 (0.0033) | 2.7080 (0.0950) | 0.9643 (0.0051) |
| | 1-Year Window | 3.9949 (0.0336) | 0.9263 (0.0085) | 2.8535 (0.1148) | 0.9806 (0.0046) | 2.7788 (0.1100) | 0.9683 (0.0081) |
| MAG | Cumulative Years | 3.5003 (0.1458) | 0.9455 (0.0081) | 2.7734 (0.1971) | 0.9816 (0.0055) | 2.7067 (0.1496) | 0.9723 (0.0029) |
| | 5-Year Window | 3.5391 (0.1628) | 0.9553 (0.0186) | 2.7503 (0.2232) | 0.9815 (0.0052) | 2.7051 (0.1602) | 0.9753 (0.0045) |
| | 1-Year Window | 3.7888 (0.1701) | 0.9546 (0.0200) | 2.8208 (0.2753) | 0.9790 (0.0049) | 2.7601 (0.2072) | 0.9778 (0.0052) |
| DBLP | Cumulative Years | 3.5319 (0.1929) | 0.9363 (0.0191) | 3.0056 (0.3967) | 0.9662 (0.0239) | 2.8149 (0.2681) | 0.9732 (0.0089) |
| | 5-Year Window | 3.6529 (0.3154) | 0.9395 (0.0230) | 2.9933 (0.4817) | 0.9708 (0.0204) | 2.7905 (0.3126) | 0.9784 (0.0044) |
| | 1-Year Window | 3.9452 (0.3568) | 0.9405 (0.0182) | 3.1159 (0.6281) | 0.9687 (0.0211) | 2.8693 (0.4099) | 0.9774 (0.0044) |

Random Selection

The aforesaid similarity in power-law fit test results for cumulative-year and 5- & 1-year sliding window analyses indicates that power-law distributions may be observed without much over-time accumulation of degrees resulting from, e.g., preferential attachment. To test this idea, publication records were randomly selected to form approximately 10% of all publications in each data. Then, subsets ranging from 10% to 100% of all records in the sampled data were randomly selected with increments of 10,000 records. Out of 9.6M MEDLINE records, for example, a starting sample of 1M records was randomly chosen. From the sampled data, a total of 91 subsets ranging from 100,000 (1%) to 1M (100%) were generated by random selection (except the case of 100%). Next, the CDF-LS power-law fit test was conducted on the degree distribution from each subset's collaboration network per disambiguation method. On the starting sample data (i.e., 1M records from MEDLINE), this process was repeated 10 times and the resulting scaling parameters and R-squared values were averaged across the same-sized subsets[12].

The results are visualized in Figure 7, where unlike Figure 4 ~ 6, each data point represents mean $\alpha$ (x-axes) and mean $R^2$ (y-axes). The arrow-headed lines represent the direction of subset size increase: the tails represent the smaller sizes and the arrow-heads larger ones.

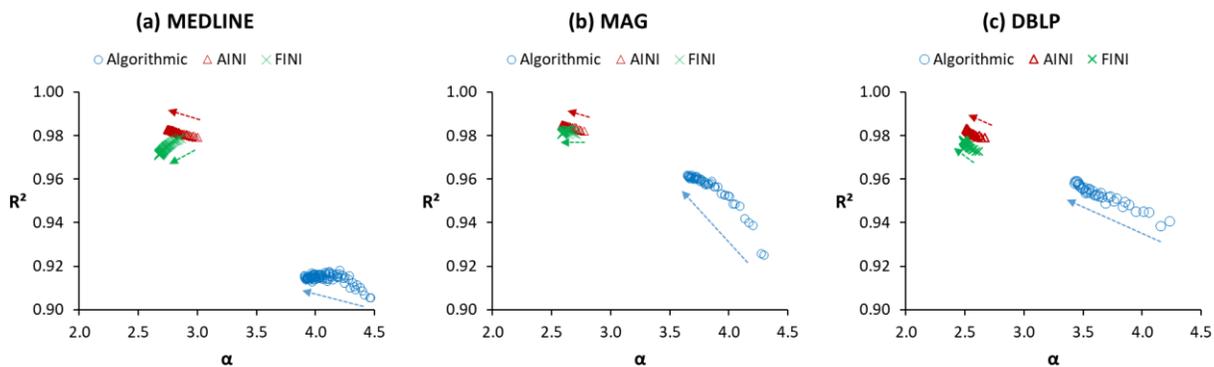

Figure 7: Trends of Scaling Parameter ($\alpha$) and R-squared Fit ($R^2$) per Disambiguation Method for Random Samples: arrow-headed lines represent the subset size increase from smaller to larger ones

In Figure 7, data points from initial-based disambiguation (red triangles and green crosses) are densely clustered in the upper-left corners. Data points for AINI are little higher than those for FINI. Commonly, they are congregated around α = 2.5 ~ 3.0 and $R^2$ = 0.97 ~ 0.99. This indicates that if relied on the initial-based name disambiguation, many collaboration networks generated from various-sized random publications can be regarded to be scale-free. In contrast, data points from algorithm-based disambiguation (blue circles) are found in the lower-right corners, stretching diagonally. This indicates that as the subset size increases from 1% to 100% of each sampled data, algorithmic disambiguation produced degree distributions getting closer to those in scale-free networks but with higher $\alpha$ and lower $R^2$ than those by initial-based disambiguation.

Considering that the test subsets were randomly selected to be 1% to at most 10% of 3.4M ~ 9.6M publication records spanning over 19 to 26 years, the power-law distributions by initial-based disambiguation are unlikely to emerge by degree accumulation from social interaction over years. Based on the disambiguation accuracy reported in Table 2 where names disambiguated by FINI and AINI were merging-prone, the power-law fit distributions seem to be shaped possibly by artefactual nodal entities that happen to combine degrees of multiple authors and thereby contribute to the formation of scale-free-like network structure. In the same vein, the power-law-like distributions by algorithmic disambiguation may also be affected, to some degree, by merged or split entities which algorithms failed to disambiguate correctly because algorithmic disambiguation also merged or split author entities even at much lower levels than initial-based disambiguation.

Error Simulation

To better understand how disambiguation errors can affect the changes of $\alpha$ and $R^2$, merging and splitting errors were simulated on the random datasets used above. Specifically, given a random dataset, a list of distinct authors disambiguated by algorithms was made. From the list, authors who will be merged into others if their names are disambiguated by AINI were selected. Then, 1% to 100% of such merging-prone authors by AINI were randomly selected and their associated name instances in the random dataset were changed into the AINI format. This randomization of merging errors was repeated on the same dataset for FINI. In contrast, splitting simulation was conducted on the list of distinct authors disambiguated by algorithms but appearing in two or more publication records in the random dataset. After randomly selecting 1% to 100% of all authors in the list, name instances of those selected authors were changed into different entities by adding unique numbers to the name instances.

---
[12] For MAG, subsets ranged from 50,000 to 500,000 records, while for DBLP, from 40,000 to 400,000 records.

In Figure 8, data points represent $\alpha$ (x-axes) and $R^2$ (y-axes) calculated for degree distributions of the random data per merging (red triangles by AINI and green crosses by FINI) or splitting (blue circles) error level. The errors increased from 1% to 100% with increments of 1%, which is denoted by the arrow-headed lines depicting the increase of error ratios: the tails represent lower ratios and the arrow-heads higher ratios[13].

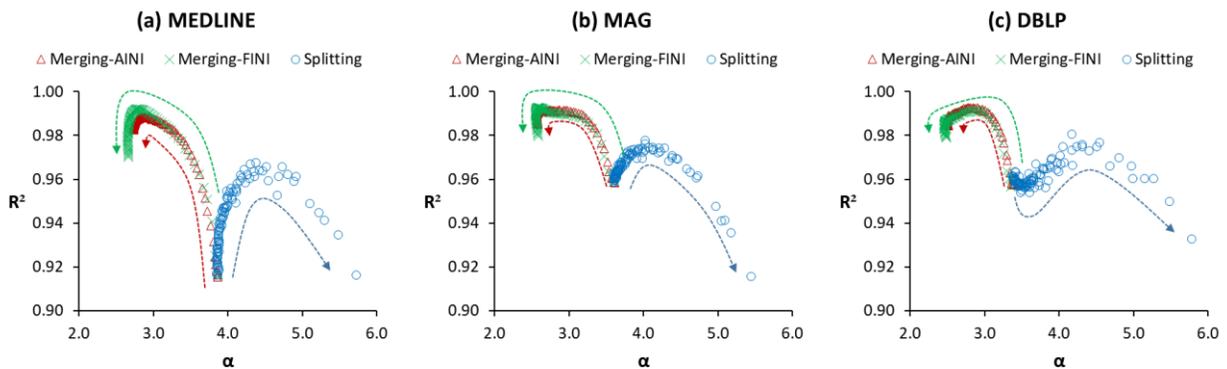

Figure 8: Trends of Scaling Parameter ($\alpha$) and R-squared Fit ($R^2$) per Disambiguation Error Ratio for Random Samples: arrow-headed lines represent the error ratio increase from lower to higher ones

A common observation across datasets is that as merging errors increased, degree distributions moved toward the upper-left corners in the figure, producing higher $R^2$ and lower $\alpha$. But the moving trends were not linear. The $R^2$ in Figure 8 increased quickly as merging error ratios increased from bottom lines (= 0%) and reached their peak, $R^2 \approx 0.99$, especially when the merging error ratios by AINI and FINI were around 40~50% in MEDLINE, 40~60% in MAG, and 20~30% in DBLP, with $2.5 < \alpha < 3.0$. After the $R^2$ peaks, data points fell vertically: degree distributions were fitted to power-law slopes with decreased $R^2$ around $\alpha = 2.5$. In contrast, splitting errors pushed blue circles to the upper-right side for many error ratios, increasing both $\alpha$ and $R^2$. After reaching $R^2 = 0.97~0.98$, the blue circles continued to fall diagonally.

According to the simulation results, the effects of merging and splitting errors on degree distributions worked in different directions for $\alpha$: merging tended to reduce it while splitting increased it. In contrast, $R^2$ showed rise-and-drop patterns as more errors, whether they are merging or splitting, were introduced to random data. However, considering that the reversed simulation of splitting corresponds to merging (i.e., blue circles moving backward), the simulation results of splitting also corroborates that merging tends to produce higher $R^2$ with lower $\alpha$, with the rise-and-drop pattern shown for the effects of merging by initial-based disambiguation.

---

[13] For visual simplicity, data points positioned outside $\alpha > 6.0$ or $R^2 < 0.90$ were excluded from visualization in each subfigure. Four data points were excluded in MEDLINE, six in MAG, three in DBLP. They were all splitting cases.

These observations imply that depending on the levels of name disambiguation errors, the same data can produce degree distributions that have different power-law slopes and fits. Especially, merging errors induced by initial based disambiguation were shown to generate degree distributions that look closer to power-law slopes with higher $R^2$ than those by algorithmic disambiguation. This may explain why merging-prone AINI and FINI produced degree distributions getting closer to power-law slopes over cumulative years and sliding windows of 5 and 1 year in Figure 4 ~ 6. In other words, the movement patterns of data points by AINI and FINI in Figure 4 ~ 6 resemble those by the simulated effects of merging in Figure 8, implying that emergence of power-law like distributions may be heavily affected by the artefactual entities merged by initial-based disambiguation.

To check this scenario, top five authors with high degrees per disambiguation method in each data were manually checked for their identities using full name, coauthor name, email address, and affiliation information, if available. The high-degree authors identified by AINI and FINI in Table 5 were found to be mixtures of multiple distinct authors with Chinese and Korean names (Kim & Diesner, 2016; Milojević, 2010a; Strotmann & Zhao, 2012). Algorithmic disambiguation also amalgamated distinct authors in MAG and DBLP although the numbers of fused authors were smaller than those by initial-based disambiguation. This implies that datasets disambiguated algorithmically are also prone to merging errors and can lead to generation of distorted degree distributions.

*Table 5: Examples of Names and Degrees (in parentheses) of Highly Collaborative Authors Per Disambiguation Method*

| Data | MEDLINE | | | MAG | | | DBLP | | |
|---|---|---|---|---|---|---|---|---|---|
| Disam-biguated by | Algorithm | AINI | FINI | Algorithm | AINI | FINI | Algorithm | AINI | FINI |
| 1 | Shizuo_A (2,508) | Wang_Y (25,990) | Lee_J (25,006) | Vu (2,829) | Wang_Y (17,014) | Wang_Y (15,914) | Li_Wei (1,763) | Wang_Y (8,594) | Wang_Y (8,381) |
| 2 | Copeland_O (2,158) | Zhang_Y (22,155) | Wang_Y (23,896) | Wang_Jun (2,788) | Zhang_Y (14,497) | Lee_J (14,215) | Wang_Wei (1,716) | Zhang_Y (8,118) | Zhang_Y (7,430) |
| 3 | Jenkins_N (2,099) | Wang_J (20,574) | Wang_J (22,049) | Wang (2,586) | Wang_J (14,131) | Wang_J (14,109) | Li_Jing (1,579) | Li_Y (7,045) | Chen_Y (7,299) |
| 4 | De Clercq_E (2051) | Li_Y (20,304) | Lee_S (21,592) | Wang_Jian (2,471) | Chen_Y (13,262) | Lee_S (13,870) | Zhang_Li (1,522) | Wang_J (6,881) | Wang_J (6,903) |
| 5 | Li_N (1,982) | Li_J (18,184) | Zhang_Y (19,386) | Li (2,398) | Liu_Y (13,124) | Zhang_Y (13,257) | Wang_lei (1,551) | Liu_Y (6,787) | Li_Y (6,614) |

(Degree Ranking on left side of table)

## Conclusion and Discussion

This study illustrates that name ambiguity can contribute to the emergence of likely scale-free collaboration networks mainly by merging author entities when author names are improperly disambiguated. Such tendency was consistently observed across three large-scale datasets representing different scientific domains. Power-law fit test with various data slicing techniques – cumulative years, 5- & 1-year sliding window, and random sampling – resulted in the same finding: initial-based disambiguation tended to generate degree distributions closer (in terms of R-squared goodness-of-fit) to power-law slopes than those created by algorithmic disambiguation. The all-initials-based disambiguation commonly used in many scale-free collaboration network studies produced the best power-law fitting results than the first-initial-based method as well as algorithm-based disambiguation. Even algorithmic disambiguation was not free of disambiguation errors and thus prone to distortion of degree distribution.

Other than the cautionary message that author name ambiguity can affect the degree distribution shape of seemingly scale-free-like collaboration networks, this study does *not* suggest that scale-free collaboration networks in prior research relying on initial-based disambiguation are results of artifacts or need to be re-examined. The main reason is that collaboration networks in many studies have been constructed under several constraints at the time of their studies. Most author names in papers before mid-2000s were recorded in the format of a forename initial(s) and a full surname(s) and lacked auxiliary information such as affiliation, which can degrade the performance of algorithmic disambiguation. In addition, in the absence of user-friendly name disambiguation packages or toolkits, the implementation of sophisticated disambiguation algorithms must have been a daunting task to many collaboration network scholars who are not adept at it. For bibliographic data obtained under these conditions, initial-based disambiguation would be the optimal solution to resolving author name ambiguity.

Another bound of this study is that the results shown in this paper cannot corroborate or dispute the prevalence of scale-free collaboration networks. This is mainly because detecting a power-law distribution in networks can be a matter of "the eye of the beholder." Some scale-free collaboration network studies do not report goodness-of-fit for their power-law test results other than visually confirming the straight-line-ness of a test distribution. In addition, any specific level of $R^2$ and other goodness-of-fit has not been agreed by scholars for judging an eligible power-law distribution. Furthermore, while part of any degree distribution can be fitted to a power-law with near perfect fit (Clauset et al., 2009), there is no consensus on how many data points in a degree distribution should be governed by a power-law regime to be "legitimately" scale-free (Stumpf & Porter, 2012). Under these practices, the validity of detection of or the universality of scale-free collaboration networks cannot be properly debated even with rigorous statistical fitting methods[14]. Thus, the findings of this study should not be accepted as evidence for or against prior studies (i.e., "is scale-free or not") but as a showcase of the degree distribution changes in collaboration networks under different name ambiguity control settings (i.e., "looks more scale-free or less").

Based on the results of this study, a few suggestions are worth noting to improve the research practice in search of scale-free collaboration networks. First, scholars should be warned that author name ambiguity can be detrimental to the study of collaboration networks by generating merged and/or split nodal entities. Recently, such distortive effects of ambiguous bibliographic data have been discussed for bibliometrics in general as well as network measures (e.g., Schulz, 2016; Strotmann & Zhao, 2012; van den Besselaar & Sandström, 2016). Beyond the evolution of collaboration networks, the fact that author name ambiguity can inflate or deflate the number of authors can affect findings of research on the growth of scientific workforce (e.g., Bebber et al., 2014; Viana, Amancio, & da Fontoura Costa, 2013), author-level analysis of citation impact such as h-index and co-citation networks (e.g., Amancio, Oliveira, & da Fontoura Costa, 2012b; Ding, Yan, Frazho, & Caverlee, 2009). So, a sensitivity test with different name disambiguation methods would be recommended for future studies before claiming detection of an author-based topological property from bibliographic data. Also, in-depth studies should follow on how the interplay of merging and splitting errors affects network structure and what levels of disambiguation errors are acceptable under what conditions for claiming knowledge discovery from ambiguous bibliographic data. Second, some collaboration network scholars who had used initial-based disambiguation began to implement algorithmic disambiguation (e.g., Martin, Ball, Karrer, & Newman, 2013; Sinatra, Wang, Deville, Song, & Barabási, 2016). In line of these efforts, researchers who plan to mine ambiguous collaboration network data may consider working together with computer and information scientists who have developed high-performing disambiguation models based on various feature engineering techniques and algorithms (e.g., Amancio, Oliveira, & da Fontoura Costa, 2012a;

---

[14] For critical discussions of fitting methods in Clauset et. al (2009), see Milojević (2010b) as well as Barabási (2018) in response to Broido and Clauset (2018)

Torvik & Smalheiser, 2009) or use bibliographic data such as Author-ity (Torvik & Smalheiser, 2009) and DBLP (Ley, 2002) where name ambiguity is controlled with high accuracy. Lastly, beyond collaboration networks, data quality can matter for networks where nodes are prone to merging or splitting errors due to ambiguous entities as in the movie actor co-appearance network (Barabási & Albert, 1999) or gene-protein interaction networks (Aladağ & Erten, 2013). This study will be a benchmark for future efforts to investigate these problems.

Acknowledgement

This work was supported by the Beta Phi Mu Eugene Garfield Doctoral Dissertation Fellowship, and grants from the National Science Foundation (grants #1561687 and #1535370), the Alfred P. Sloan Foundation and the Ewing Marion Kauffman Foundation. I would like to thank Jana Diesner, Catherine L. Blake, and Vetle I. Torvik at the University of Illinois at Urbana-Champaign, Michelle Shumate at Northwestern University, and Seok-Hyoung Lee at KISTI. I am also thankful to anonymous reviewers for their insightful comments.

Appendix A: Power-law fitting on data filtered for hyper-authorship

Hyper-authorship can blur our understanding of evolving collaboration networks because it makes transient (i.e., publishing a single paper) authors super-nodes with extremely high degrees. So, some scholars studying scale-free collaboration networks have excluded papers with hyper-authorship. Choosing a specific author-size per paper can be arbitrary because there is no agreed number of authors per paper for defining hyper-authorship. As this paper analyzed datasets representing three different fields, the criterion of hyper-authorship was hard to decide. Thus, top 1% papers high in the number of authors per paper were excluded from analysis in this paper, as visualized in Figure 9.

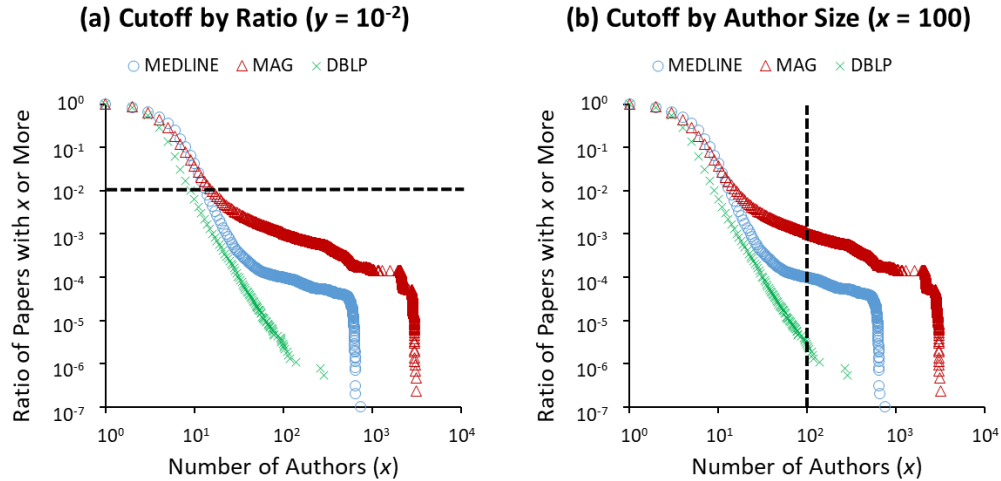

Figure 9: Comparison of Paper Exclusion Thresholds by (a) Ratio and (b) Hyper-authorship

In the figure, the cumulative distribution of the number of authors per paper is plotted on doubly logarithmic scales for whole data. The left subfigure represents this study's decision: top 1% ($10^{-2}$) of papers that have large number of authors, resulting in the exclusion of papers with x ≥ 13 (MEDLINE), x ≥ 16 (MAG), and x ≥ 9 (DBLP). The right subfigure shows a cutoff decision based on a specific number (x > 100) of authors per paper. Note that this cutoff led to widely different ratios (*y*-axes) of excluded papers in each dataset. Figure 10 reports the fitting results repeated on cumulative, 5-year, and 1-year data in which papers with more than 100 authors were excluded.

Overall, changing patterns of $\alpha$ and $R^2$ per disambiguation method for MEDLINE, MAG, and DBLP are quite similar to those reported in Figure 4 ~ 6: $\alpha$ by AINI and FINI falls mostly within 2 < α < 3 scoring higher $R^2$ values than those by Algorithmic. Note that the R-squared values show larger variations than in Figure 4 ~ 6: vertically stretched in MAG and DBLP. Distributions by algorithmic disambiguation also show the similar patterns in Figure 4 ~ 6 except that $R^2$ values are lower spreading below 0.90 in MAG and DBLP. These wide variations of $R^2$ might be because the inclusion of papers with author size <= 100 and author size > 16 (MAG) and 9 (DBLP) added nodes with large degrees, which makes the shape of CDF-based degree distribution less smooth or with more curvature, leading to lower $R^2$.

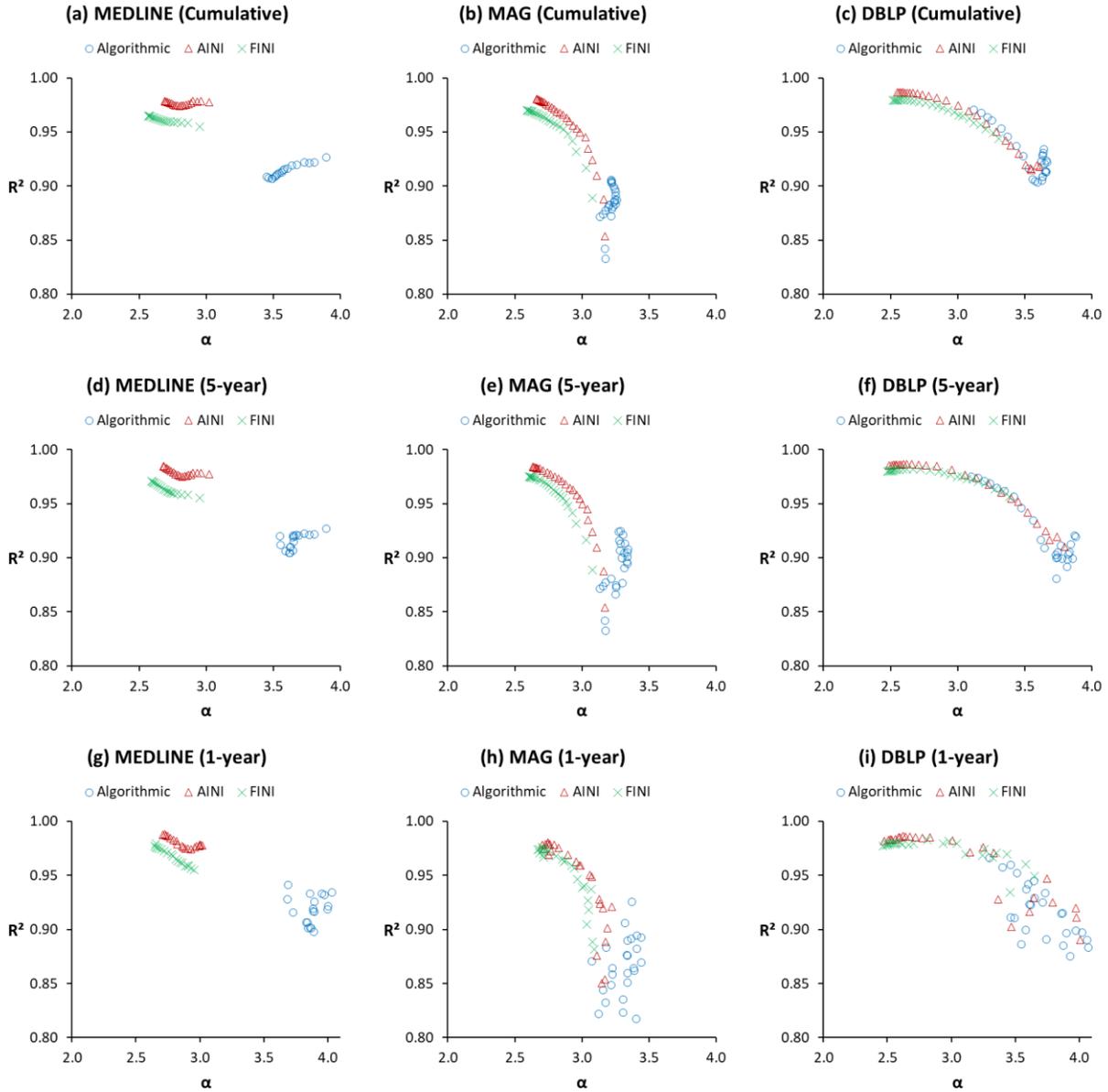

Figure 10: Trends of Scaling Parameter (α) and R-squared Fit ($R^2$) Per Disambiguation Method over Cumulative, 5- & 1-year Window with Hyper-authorship Papers Excluded

Appendix B: Power-law fitting on limited *x* values

This section shows how the selection of *x* values to fit a power-law affects the findings of this study. For this, from the same datasets used for Figure 4 ~ 6, minimum *x* (*min*) values were decided by the maximum likelihood estimation with KS statistics described in Clauset et al. (2009) using an R package poweRlaw (https://cran.r-project.org/web/packages/poweRlaw/index.html). Then, power-law fitting was conducted on the *x* values equal to or greater than *min*. Figure 11 ~ 13 report the results for cumulative, 5-year, and 1-year window data each.

According to the figures, the findings in Figure 4 ~ 6 are confirmed by the power-law fitting tested on specific *x* value ranges: (1) initial-based disambiguation produced power-law slopes approaching and falling within the canonical $2 < \alpha < 3$ with high $R^2$ under various data slicing methods and (2) algorithmic disambiguation produced decreasing α but stretched beyond α > 4 (steeper slope), which was rare in Figure 4 ~ 6. This means that power slopes were fitted on *x* values in the tails of degree distributions with downward curvature. Another difference is that the algorithmic disambiguation method generated α with higher $R^2$ than when fitted on all *x* values. This is, however, not unexpected because the *x* value ranges are optimized by selecting *min* that is supposed to generate the best straight line. Also note that the ratios of fitted *x* values over all *x* values are very low: especially, below 1% for algorithmically disambiguation degree distributions. This means that power-law fitted by x-minimum calculation can describe extremely small portion of network actors.

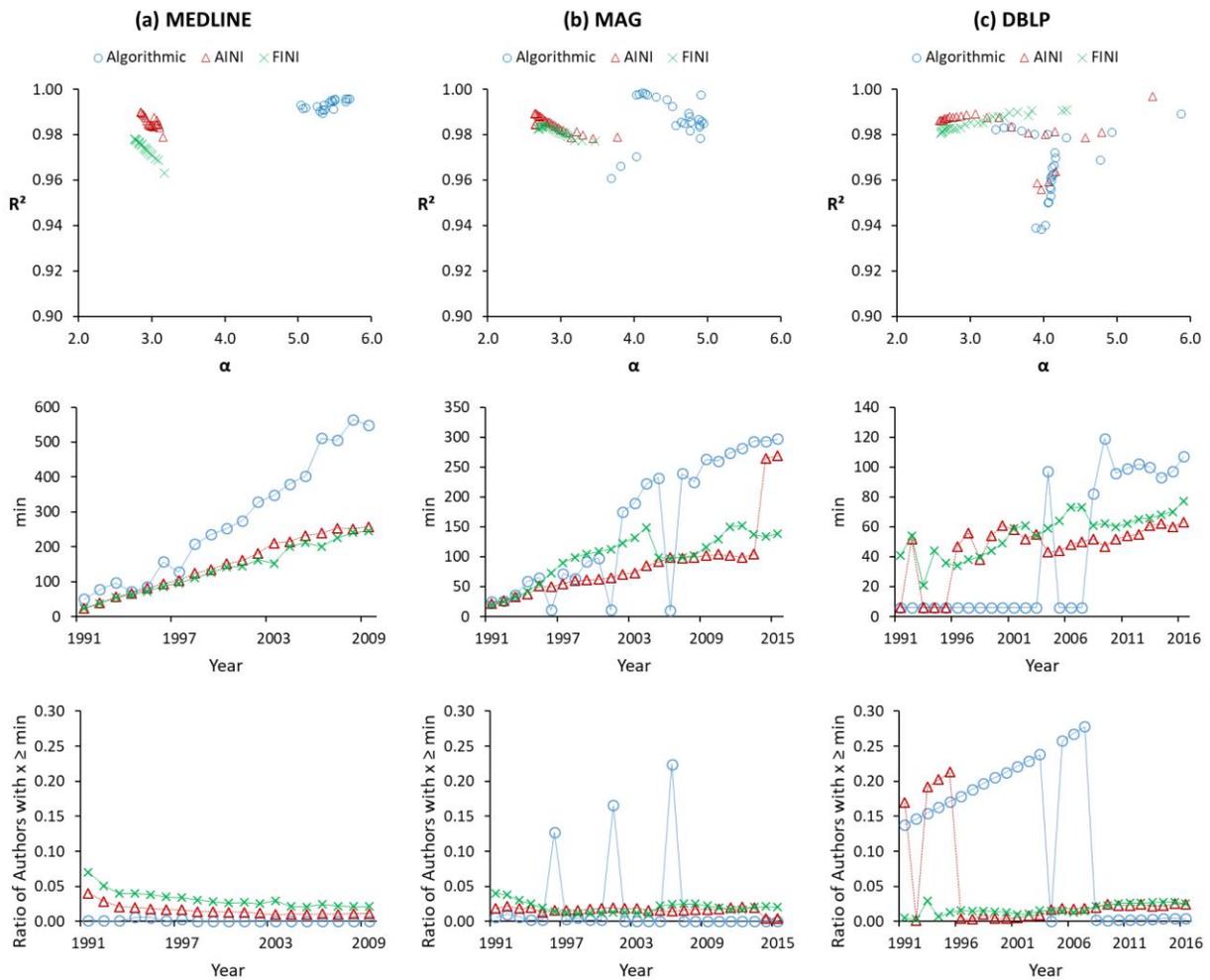

*Figure 11: Trends of Scaling Parameter (α) and R-squared Fit ($R^2$) Per Disambiguation Method over Cumulative Years Tested on Limited x Values (upper), Changes of Minimum x Values (min) over Cumulative Years (middle), and Ratios of Authors with x ≥ min over Cumulative Years (lower)*

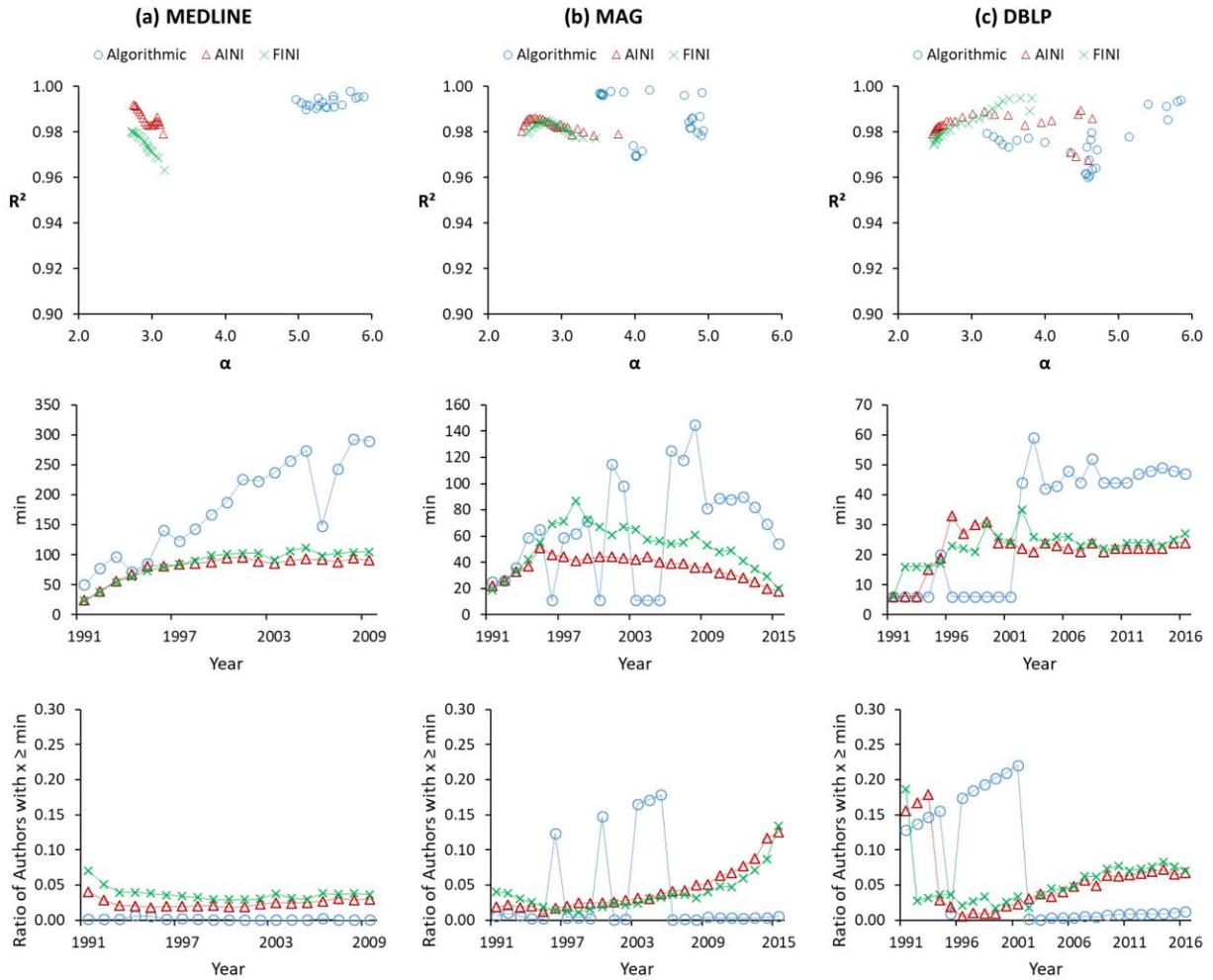

Figure 12: Trends of Scaling Parameter (α) and R-squared Fit ($R^2$) Per Disambiguation Method over 5-Year Window Tested on Limited x Values (upper), Changes of Minimum x Values (min) over 5-Year (middle), and Ratios of Authors with x ≥ min over 5-Year (lower)

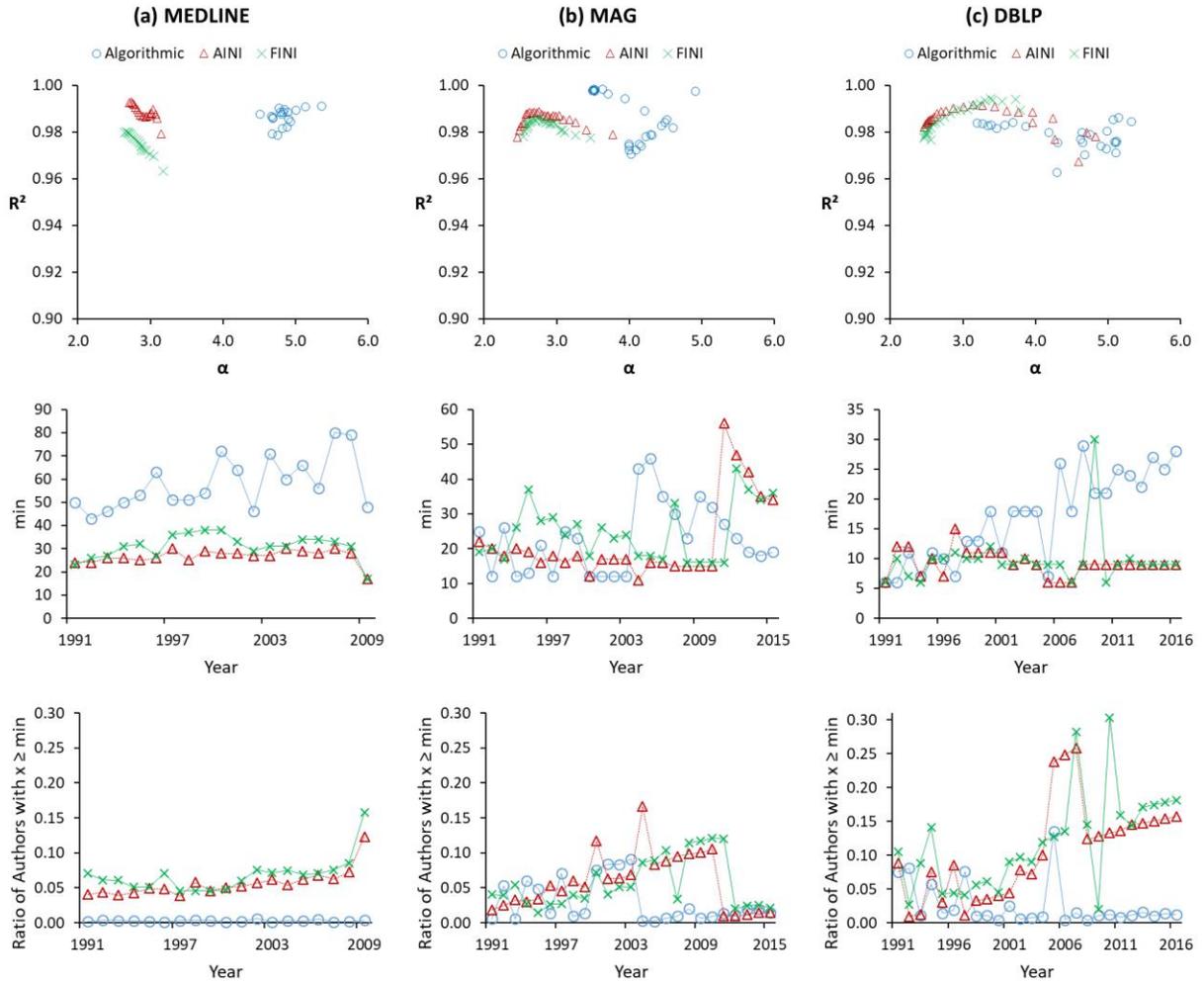

Figure 13: Trends of Scaling Parameter (α) and R-squared Fit ($R^2$) Per Disambiguation Method over 1-Year Window Tested on Limited x Values (upper), Changes of Minimum x Values (min) over 1-Year (middle), and Ratios of Authors with x ≥ min over 1-Year (lower)